\documentstyle{article}
\begin{document}
\twocolumn[
\hrule
\vspace{1mm}
Presented by Z.W\l odarczyk at the {\it IX$^{th}$ Int. Symp. 
on Very High Energy Cosmic Ray Int.}, Karlsruhe, Germany,
August 19-23, 1996//
\vspace{1mm}
\hrule
\vspace{3mm}
\noindent
{\large \bf Centauro as Probe of Deeply Penetrating Component in Cosmic
Rays}\\

\noindent
G. Wilk$^a$ and 
Z.W\l odarczyk$^b$

\noindent
$^a$Soltan Institute for Nuclear Studies, Nuclear Theory Department, 
Warsaw, Poland\\
(e-mail: wilk@fuw.edu.pl)\\

\noindent
$^b$Institute of Physics, Pedagogical University, Kielce, Poland\\
(e-mail:wspfiz@sabat.tu.kielce.pl)\\

The requirements for observing {\it Centauro}-like phenomena and
their role as possible probes of deeply penetrating component in
Cosmic Rays are discussed.\\
\vspace{5mm}
]
\noindent
\section{Introduction}
The {\it Centauro} and {\it mini-Centauro} events, characterized by
the extreme imbalance between hadronic and gamma-ray components among the
produced secondaries \cite{L}, are the best known examples of
numerous unusual events reported in Cosmic Ray (CR) experiments \cite{ZW}.
There are numerous attempts to  explain them as: $(i)$ different
types of isospin fluctuations or formation of disoriented chiral
condensate (DCC) \cite{ACMAB}; $(ii)$ multiparticle Bose-Einste\-in
correlations \cite{P}; $(iii)$ strange quark matter formation or
interaction \cite{SQM}. All of them reproduce many features of {\it
Centauros} in a single collision 
but fail to explain the substantial number of
interactions contributing to the development of families observed at
mountain altitudes among which {\it Centauro} were observed.  

We demonstrate that families produced at mo\-un\-tain altitudes are
{\it insensitive} to any isospin fluctuations mentioned above (cf.
\cite{FLUC}). This means that {\it Centauro} must originate from some
very penetrating projectiles. We discuss two  scenarios leading to
such enhanced penetrability: fluctuations of elementary cross sections
\cite{FCS} and propagation of chunks of Strange Quark Matter (SQM)
({\it stran\-ge\-lets}) in the atmosphere \cite{STR}. 

\section{Centauros in atmospheric cascades }
Isospin fluctuations is characterized by the ratio of
neutral to total pion production yields,
\begin{equation}
r \, =\, \frac{N_{\pi^0}}{N_{\pi^0} + N_{\pi^{ch}}}\, =\, 
        \frac{N_{\pi^0}}{N} ,  \label{eq:R}
\end{equation}
where $N$ denotes total pion multiplicity. If isotopic spin is to be 
conserved $r$ should be distributed according to binomial
distribution (BD) (which for large $N$ is well approximated by a 
gaussian with width $\sigma = 1/\sqrt{3N}$) widely used in most Monte
Carlo generators. However, models \cite{ACMAB,P} predict a strong
deviations from simple BD formula, namely for large $N$ one has
\begin{equation}
P(r)\, =\, \frac{1}{2\sqrt{r}}\, , \label{eq:DCC}
 \end{equation}
i.e., $r$ differs substantially from the naive value of $\langle  
r\rangle = 1/3$. The natural question therefore arises: does this
fact influence also characteristics of atmospheric families observed
in the same experiments? To answer this question we have calculated
(using standard RR-Y00 model from SHOWERSIM software \cite{MC})
distributions $P(\varepsilon)$ of the energy fraction   
\begin{equation}
\varepsilon \, =\, \frac{\sum E_{\gamma}}{\sum E_{\gamma}\, +\,
\sum E_{\gamma}^h} \label{eq:E}
\end{equation}
of electromagnetic $(\sum E_{\gamma})$ and hadronic
$(\sum E_h)$ components of gamma-hadron families (with total
visible energy $\sum E_{\gamma} + \sum E_{\gamma}^h \geq 100$ TeV)
re\-cor\-ded in emulsion chambers at mountain altitudes. The results are
shown in Fig. 1 where $P(\varepsilon)$ for DCC type of
fluctuations (\ref{eq:DCC}) essentially coincides with the standard
behaviour of families described by BD distribution. It means that all
scenarios of isospin fluctuations lead to essentially {\it the 
same} families at mountain altitudes. (We have checked it
for $P(r)$ ranging from the pure {\it Centauro-Anticentauro} 
production case corresponding to $P(r) = \frac{1}{3} \delta(r - 1) +
\frac{2}{3} \delta(r)$ to the broadest smooth distribution of the
form $P(r)= 2(1 - r)$; the same conclusion holds also for
distributions of the corresponding multiplicity fractions $P(\eta)$,
$\eta =N_{\gamma}/(N_{\gamma} + N_h)$.). It means that {\it
Centauros} need projectiles of the unusually large penetrability in
the atmosphere. We shall discuss now two possible scenarios 
leading to such a large penetrability: (i) fluctuation of
hadronic cross sections \cite{FCS} and (ii) propagation of
strangelets in the atmosphere \cite{STR}.

\begin{figure}[h]
\setlength{\unitlength}{1cm}
\begin{picture}(7.28,7.11)
\includegraphics{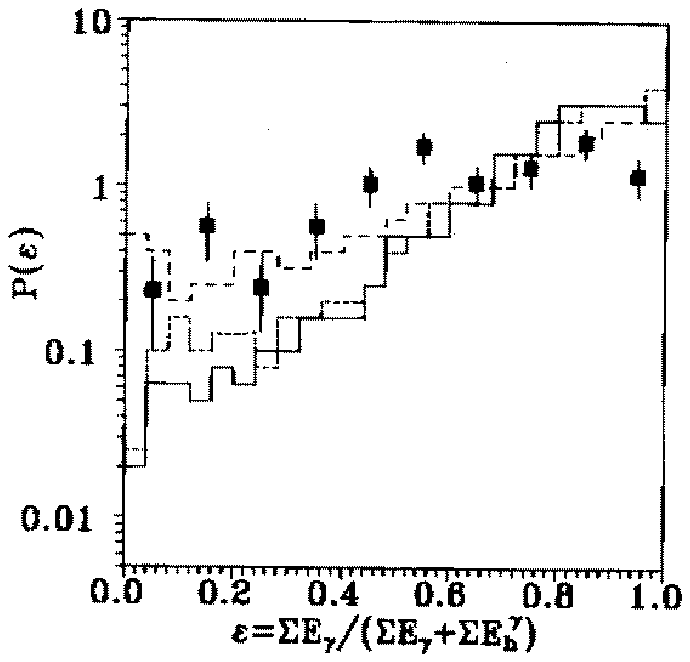}
\end{picture}
\vspace{-0.5cm}
\end{figure}
\begin{minipage}[h]{6.8cm}
\noindent
Figure 1. Distributions $P(\varepsilon)$ for gamma-hadron families
detected at mountain altitudes. Results for the standard BD
distribution (solid line) and DCC one as given by eq.(\ref{eq:DCC})
(without CSF - dotted line and including CSF - dashed line) are
compared with data from $135$ Chacaltaya families \cite{CH}
(squares).  
\end{minipage}
%\vspace{2mm}

\subsection{Fluctuations of hadronic cross sections}
Origin of the cross section fluctuations (CSF) is traced down to the
compositeness of hadrons and their extension. Depending on the
temporal and space configurations of partons (quarks and gluons)
composing both colliding hadrons one can have different interaction
cross sections for a given interaction with the mean value being
equal to $\sigma_{hh}$. In \cite{FCS} CSF allowed to explain the so
called {\it long flying component} of CR, in accelerator data they
manifest themselves in the diffraction dissociation events. The
significance of the CSF in  present situation is illustrated in Fig.
1 where out of the two $P(\varepsilon)$ distributions of DCC type
(\ref{eq:DCC}) only that with CSF (simulating deeply penetrating
component of CR) can describe experimental data (using cross section
fluctuation characterized by $\omega = \left(\langle \sigma^2\rangle 
- \langle \sigma\rangle^2\right)/\langle \sigma \rangle^2 = 0.2$
\cite{FLUC}). 

\subsection{Strangelets in the atmosphere}
In \cite{STR} we have proposed scenario of the extraterrestial
strangelets penetrating deeply in the atmosphere in which heavy lumps
of strange quark matter of mass number $A_0 \sim 1000$ arrive at the
top of atmosphere and degradate in the successive interactions with
air nuclei until reaching some critical value of $A_{crit} \sim 320$.
Below this value they simply desintegrate into nucleons \cite{STR}.
{\it Centauro} events \cite{L} interpreted in terms of this approach would
correspond to $A_0 = 2000$ if detected at Chacaltaya ($540$ g/cm$^2$)
and to $A_0 = 2350$ if detected at Pamir altitudes ($600$ g/cm$^2$).
For the mass spectrum $N(A_0) \cong \exp( - A_0/130)$ discussed in
\cite{STR} the above mass numbers correspond to the observed flux ratio of
{\it Centauro} events, $N_{Pamir}/N_{Chacaltaya} \cong 0.066$ explaining
nicely the small number of {\it Centauros} detected at Pamir
altitude. The observed $5$ events on the area $200$ m$^2$yr and $1$  
event on the area $600$ m$^2$yr from the Chacaltaya and Pamir
experiments respectively, \cite{CP}, lead to the observed flux ratio
equal to $N_{Pamir}/N_{Chacaltaya} = 0.06$. 

\begin{figure}[h]
\setlength{\unitlength}{1cm}
\begin{picture}(7.28,6.92)
\includegraphics{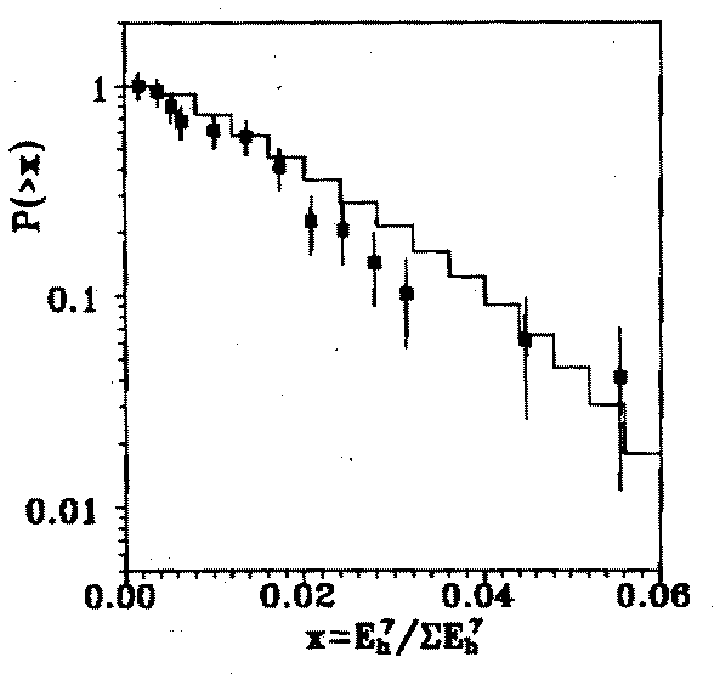}
\end{picture}
\vspace{-0.3cm}
\end{figure}
\begin{minipage}[h]{6.8cm}
\noindent
Figure 2. Integral distribution of fractional energy of hadrons
in {\it Centauro} event \cite{L} (squares) compared with our predictions
for $E_h = $const and fluctuating partial inelasticity $k_{\gamma}$
distributed according to UA5 Monte Carlo algorithm for nuclear
interactions \cite{T}.
\end{minipage}
\vspace{2mm}

In this scheme {\it Centauros} are baryon-emitting events. Nucleons
evaporated from the strangelet decaying when its $A < A_{crit}$
deeply in the atmosphere are recorded in the emulsion chamber.
Despite the fact that energies $E_h$ of all evaporated baryons are
very close to each other, the observed visible energies
$E^{\gamma}_h$ can still exhibit behaviour observed experimentaly. In
Fig. 2 we show the integral distribution of fractional energy $x =
E^{\gamma}/\sum E^{\gamma}_h$ of hadrons detected in the {\it
Centauro} event \cite{L} compared to our predictions. In our
simulations we start with sampling energy of such hadrons from the 
initial distribution $N(E_h) \cong \delta(E_h - E_0)$ and transform
it into $E^{\gamma}_h = k_{\gamma}E_h$ with the help of the partial
inelasticity ($k_{\gamma}$) distribution given by
\begin{equation}
f(k_{\gamma})\, {\rm d}k_{\gamma}\, =\, \frac{1}{\Gamma(\alpha)}\, 
       \left( \frac{k_{\gamma}}{\beta} \right)^{\alpha - 1}
       \exp\left( - \frac{k_{\gamma}}{\beta} \right)\, 
       \frac{{\rm d}k_{\gamma}}{\beta} \label{eq:INEL}
\end{equation}
with $\alpha = 1.05$ and $\beta = 0.145$ obtained by using UA5 Monte
Carlo algorithm as a model of nuclear interactions \cite{T}. In such
approach the energy distribution in {\it Centauro} event reflects just the
partial inelasticity distribution.

\section{Summary and conclusions}
{\it Centauros} can occur either as pionic events or as purely
baryonic ones. In the first case they can be described by some
apparent strong isospin fluctuations. However, their occurence at
mountain altitudes indicates additional strong penetrability of the
projectile causing such event. We claim that in this case such
penetrability could be provided by the CSF mechanism, cf. Fig. 1. In
the second case we show that they can be products of strangelets
penetrating deeply into atmosphere. Both flux ratio of {\it
Centauros} produced at different experiments and energy distribution
within them can be successfully described by such concept.

\end{document}